\documentclass[fleqn,usenatbib]{mnras}

\usepackage{newtxtext,newtxmath}

\usepackage[T1]{fontenc}
\usepackage{ae,aecompl}

\usepackage{graphicx}	
\usepackage{amsmath}	
\usepackage{amssymb}	
\usepackage{caption}

\newcommand{\jed}[1]{~{\rm #1}}

\title[Time resolved spectroscopy of WD~1145+017]{Time resolved spectroscopy of 
dust and gas from extrasolar planetesimals orbiting WD~1145+017
\thanks{Based on service observations (proposal SW2015b15) made with the William Herschel Telescope
operated on the island of La Palma by the Isaac Newton Group of Telescopes 
in the Spanish Observatorio del Roque de los Muchachos 
of the Instituto de Astrof\'isica de Canarias.}}

\author[M.~Karjalainen et al.]{Marie~Karjalainen$^{1}$\thanks{E-mail: mh@ing.iac.es},
Ernst~J.W.~de~Mooij$^{2,3}$, Raine~Karjalainen$^{1}$, Neale~P.~Gibson$^{3}$ 
\\
$^{1}$Isaac Newton Group of Telescopes, Apartado de Correos 321, Santa Cruz
de La Palma, E-38700, Spain\\
$^{2}$School of Physical Sciences, and Centre for Astrophysics and Relativity, Dublin City University, Glasnevin, Dublin 9, Ireland\\
$^{3}$Astrophysics Research Centre, School of Mathematics and Physics, Queen's University Belfast, Belfast BT7 1NN, UK
}

\date{Accepted XXX. Received YYY; in original form ZZZ}

\pubyear{2018}

\begin{document}
\label{firstpage}
\pagerange{\pageref{firstpage}--\pageref{lastpage}}
\maketitle

\begin{abstract}
Multiple long and variable transits caused by dust from possibly disintegrating asteroids 
were detected in light curves of WD~1145+017. We present time-resolved spectroscopic 
observations of this target with QUCAM CCDs mounted in the Intermediate dispersion 
Spectrograph and Imaging System at the 4.2-m William Herschel
Telescope in two different spectral arms: the blue arm covering 3800--4025~\AA\,
and the red arm covering 7000--7430~\AA. When comparing individual transits in 
both arms, our observations show with $20\,\sigma$ significance 
an evident colour difference between the in- and out-of-transit data 
of the order of 0.05--0.1 mag, where transits are deeper in the red arm.
We also show with $>6\,\sigma$ significance that
spectral lines in the blue arm are shallower during transits than out-of-transit. 
For the circumstellar lines it also appears that during transits the reduction in 
absorption is larger on the red side of the spectral profiles. Our results confirm previous 
findings showing the {\it u}$^{\prime}$-band excess and a decrease in line absorption during transits. 
Both can be explained by an opaque body blocking a fraction of the gas 
disc causing the absorption, implying that the absorbing gas is between
the white dwarf and the transiting objects. Our results also demonstrate the capability of 
EMCCDs to perform high-quality time resolved spectroscopy of relatively faint targets.
\end{abstract}




\section{Introduction}\label{intro}

Most Milky Way stars are orbited by planets \citep{Cassan12}, and more than 
97 per cent are expected to end their lives as white dwarfs \citep{Althaus10}. 
However, the evolution of host stars after the main sequence will change properties of 
original planetary systems. While planets closer than approximately 5 AU will most likely 
not survive the post-main-sequence lifetime of their parent star, 
any planet with semimajor axis greater than approximately 5 AU 
will survive, and its semimajor axis will increase due to stellar mass loss
\citep{Burleigh02,Debes02}. Numerical integrations show that 
some planets in previously stable orbits around a star 
undergoing mass loss will become dynamically unstable, because the stability of adjacent 
orbits to mutual planet-planet perturbations depends on the ratio of the planet mass 
to the central star's mass \citep{Debes02}. A planetesimal whose orbit is
significantly perturbed can be ejected from the system, but in some cases
may instead get sufficiently close to the white dwarf that tidal forces will be 
capable of disrupting it \citep{Jura03,Veras14}.

The indirect evidence of such disrupting events is the presence of debris
discs around at least 4 per cent of white dwarfs \citep{Rocchetto15} and
metal lines in the spectra of around 25--50 per cent of white dwarfs 
\citep{Zuckerman10, Koester14}. Theoretical works predict that 
circularisation of the dust resulting from the disrupting event forms a 
circumstellar debris disc. Within the disc, dust grains lose angular-momentum 
and are eventually sublimated and subsequently accreted onto the white dwarf, 
resulting in the appearance of metal lines in the stellar spectrum \citep{Jura03,Debes12}.
Since the strong surface gravity of white dwarfs causes all heavy elements to settle  
below the photosphere on time-scales of days to Myrs \citep{Wyatt14}, 
`pollution' in white dwarfs is a strong indication of external ongoing accretion.
It was shown that the accreted material does not originate from the interstellar
medium \citep{Farihi10} and its composition is similar to terrestrial material in the Solar
system \citep{Zuckerman07, Klein10, Gaensicke12}. The most likely interpretation is that 
these white dwarfs are contaminated by circumstellar matter, i.e. the rocky remains 
of terrestrial planetary systems \citep{Farihi10}.

\subsection{WD~1145+017}

So far the most compelling confirmation of the remnant planetary system has been the
detection of asymmetric transits in the light curve of WD~1145+017
(hereafter WD~1145) \citep{Vanderburg15}, caused by the material from possibly disintegrating
planetesimal orbiting near the Roche limit. WD~1145 also shows an infra-red
excess characteristic for a dust disc \citep{Vanderburg15}, and metal
photospheric lines in its spectrum \citep{Xu16, Redfield17}. In addition, a
circumstellar gas was detected in WD~1145 by \citet{Xu16}, through observing 
numerous circumstellar absorption lines with line widths of approximately $300$\,km\,s$^{-1}$.
No significant magnetic field was detected \citep{Farihi18}.

Transits of WD~1145 first discovered in data from \textit{Kepler} extended mission \textit{K2}
\citep{Howell14} showed several distinct periods in the range of 4.5--4.9 h
and were variable in depth \citep{Vanderburg15}. Asymmetric transits were also
observed during the first ground-based follow up \citep{Croll17}. 
Subsequent ground-based observations
\citep{Gaensicke16,Rappaport16,Gary17,Rappaport18} showed that most of transits
are fairly symmetric and the system evolves quickly, with the overall dip activity 
increasing from 2014 (\textit{Kepler K2}) until mid-2017, with only a small activity 
decrease in mid-2016. \citet{Rappaport16} found that nearly all transits appear to
drift systematically in phase with respect to the presumed orbital period of
an asteroid. They explain this drifting motion as due to smaller fragmented bodies 
that break off from the asteroid and go into a slightly shorter-period orbit. The
fragments are heated up by the white dwarf until there is a freely escaping flow of
vapours which then condense into dust clouds. Radiation pressure from the
white dwarf may then
push some of the dust into a trailing configuration analogously to solar
system comets. They estimate that these fragments last for only a few weeks before 
disintegrating, with the maximum observed longevity of > 58 days. 
\citet{Gary17} observed fragments with lifetimes up to half a year, and
explain it using quasi-continual dust production. They also assume that
collisions are behind some of the observed transit events.
\citet{Rappaport18} reported the highest level of optical activity
of WD~1145 since the discovery, and observed some transits lasting up to 2 hours. 
This is a clear change compared to earlier observations with transit duration of 3--12 minutes \citep{Gaensicke16}. 
  
There have been several attempts to detect varying transit depth of WD~1145 in
multiwavelength observations to constrain the properties of the transiting
material. \citet{Croll17} compared the 
transit depths in the {\it V}- and {\it R}-bands, and found no significant difference,
concluding that the radius of dust particles must be equal to $0.15\,\rm{\mu m}$ or
larger, or $0.06\,\rm{\mu m}$ or smaller. 
\citet{Alonso16} observed in the wavelength range 480--920 nm, and found
no significant differences in the flux between four bands centred at 530,
620, 710 and 840 nm, from which particle sizes $\leq0.5\,\rm{\mu m}$ can be
excluded for most common minerals. \citet{Zhou16} performed simultaneous
observations in the optical and near-infrared, over the wavelength range of
0.5--1.2$\,\rm{\mu m}$, 
with no measurable difference in transit depths for multiple photometric
pass-bands, allowing them to place a lower limit of $0.8\,\rm{\mu m}$ on the grain
size in the putative transiting debris cloud. \citet{Xu18} presented
observations in the optical, {\it $K_s$} and $4.5\,\rm{\mu m}$ bands, and found
the same transit depths at all wavelengths from which they conclude that
there is a deficit of small particles (with radii $\lesssim1.5\,\rm{\mu m}$) in
the transiting material. Most recently, results from simultaneous fast optical
spectrophotometry and broad-band photometry of WD~1145 were published by
\citet{Izquierdo18}, who found no significant colour differences between five
wavelength bands over the wavelength range 4300--9200~\AA. 

Interestingly, \citet{Hallakoun17} reported the detection
of `bluing' during transits, when transits are deeper in the redder bands,
with a {\it u}$^{\prime}-${\it r}$^{\prime}$ colour difference of up 
to $-0.05$~mag.
They concluded that the observed colour difference is most likely due to
reduced circumstellar absorption in the spectrum during transits, which 
indicates that the transiting objects and the gas share the same
line-of-sight, and that the gas covers the white dwarf only partially, as
would be expected if the gas, the transiting debris, and the dust causing
the infrared excess are part of the same general disc structure. 

The suggestion of reduced circumstellar absorption during transits is further supported by  
\citet{Redfield17}, who found short-term
variability in the circumstellar gas lines coinciding with the most significant
photometric transits. They observed that the absorption on the red side of
spectral profiles decreased during transits, and concluded that gas is between 
the white dwarf and transiting objects. They also observed circumstellar
absorption variability on a scale of months, when over the course of a year 
blueshifted absorption disappeared, while redshifted absorption
systematically increased. They proposed an eccentric inclined disc that 
can explain both long and short-term variations.  
Recently, \citet{Cauley18} reported that over the course of 2.2 years
their spectra show complete velocity reversals in the circumstellar
absorption, moving from strongly redshifted in April 2015 to strongly
blueshifted in June 2017. They suggest that eccentric circumstellar gas
rings undergoing general relativistic precession with a precession period of
approximately 5.3 years can explain the velocity reversals. 
Having simultaneous spectrophotometry and photometry of WD~1145, \citet{Izquierdo18}
recently unambiguously confirmed the equivalent width decrease of the circumstellar
absorption lines during the deepest and longest transit in their
observations, supporting spatial correlation between the circumstellar gas
and dust.

To aid understanding the WD~1145 system, in this study we present time resolved 
optical spectroscopy that has sufficient spectral resolution to measure line
strengths and circumstellar gas velocities, and at the same time provides high time 
resolution to sample transits by material possibly from disintegrating planetesimals
with a typical duration of few minutes. In Sects.~\ref{Observations} and \ref{reduction} we describe our 
observations and data reduction procedure. The photometric and spectroscopic
data analysis is presented in Sects.~\ref{photometric} and
\ref{spectroscopic}, and we discuss our results in Sect.~\ref{Conclusions}.

\begin{figure*}
\includegraphics[width=\textwidth]{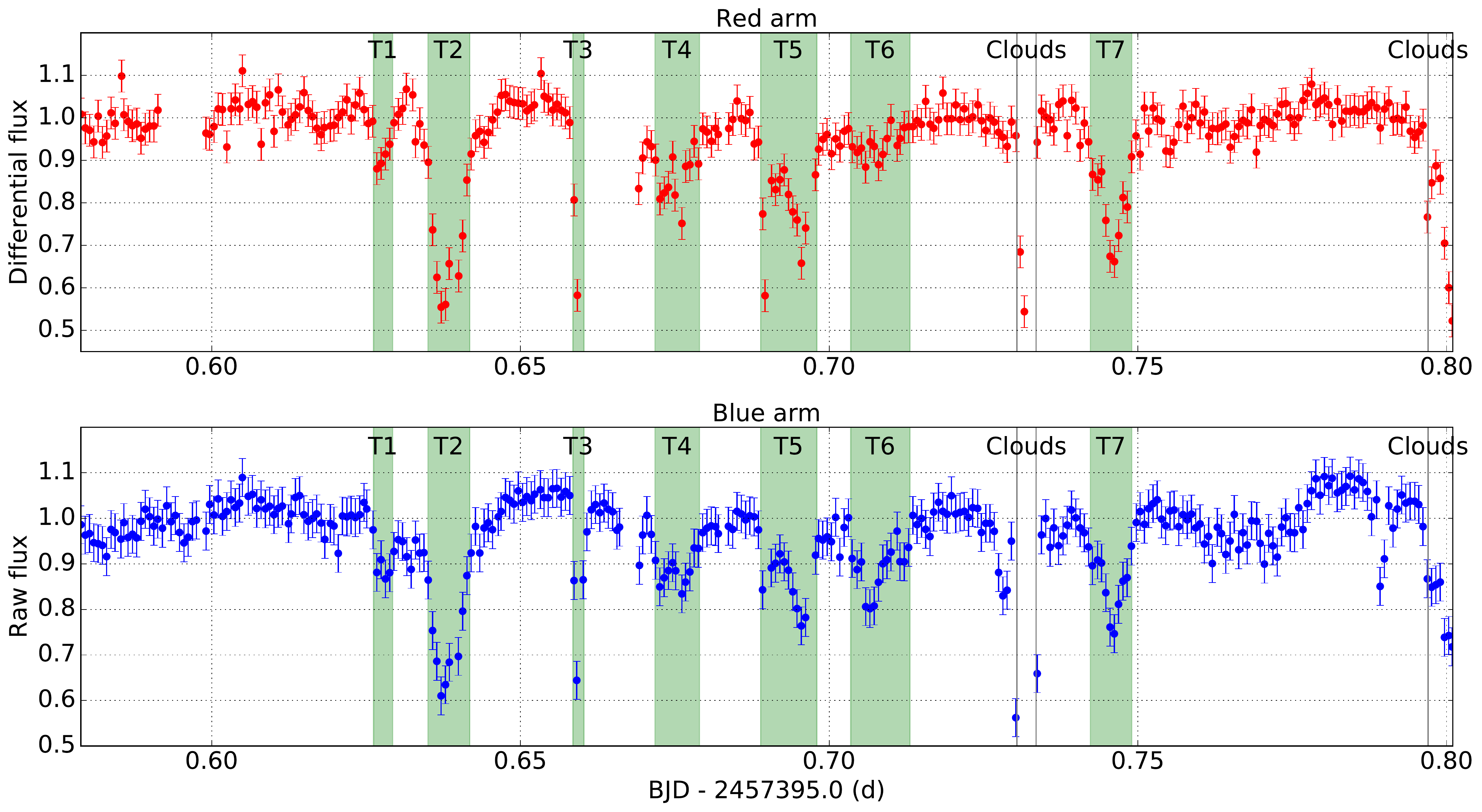}
\caption{Photometric light curve observed during the night of 2016 January 7 with
the ISIS/QUCAM2 red arm (top, red filled circles) and ISIS/QUCAM3 blue arm
(bottom, blue filled circles), binned to
an effective exposure time of 1 minute. We label individual transits T1--T7 (green
background), and also clouds passing by (between vertical lines). 
The light curves were normalized fitting a 4th order polynomial to the
out-of-transit data. Photometric uncertainties were calculated as rms errors over
out-of-transit normalized light curves, using data binned in 1 minute in time.
BJD is Barycentric Julian Date.}
\label{plot1}
\end{figure*}

\section{Observations}\label{Observations}

We observed WD~1145 on the night of 2016 January 7 with the QUCAM CCDs \citep{Tulloch11}
mounted in the Intermediate dispersion Spectrograph and Imaging System (ISIS) at the  
4.2-m William Herschel Telescope, La Palma, Spain. ISIS is a double-armed, 
medium-resolution long-slit spectrograph. Use of dichroic filters permits 
simultaneous observing in the blue and red arms. QUCAM2 and QUCAM3 are 
electron-multiplying CCD cameras in which the signal is amplified
before readout, capable of performing high-speed or faint-target spectroscopy. 
In addition, both CCDs have a frame-transfer buffer which allows on chip storage
of an image so that an exposure can be started while the previous one is still
being read out. We used fast mode with an exposure
time of 3.6\,s (the minimum readout time of full frames for QUCAMs in fast mode)
and cubes of 150 images before each read-out. The advantage of fast
mode is minimized dead-time and essentially zero readout noise. 

We used a wide (10 arcsec) slit to minimize systematic errors and slit losses, 
while also preventing a significant contribution from the sky background 
\citep[e.g.][]{Gibson13}. To better monitor our systematic errors, we used full-frame readout 
in order to observe the reference star 2MASS~11482519+0130345 simultaneously with our target. 
With the R1200B grating on the ISIS blue arm, the QUCAM3 provided 
a wavelength coverage of 225~\AA. This allowed us to cover the strong
\ion{Ca}{ii} H \& K lines, and \ion{He}{i} 3888.65~\AA~line (from 3800--4025~\AA).
With the R600R grating on the ISIS red arm, we selected a region from 7000--7430~\AA~which 
is relatively clean from telluric lines, while also offering a significant 
wavelength baseline to allow for the grain-size to be measured.

The observations took place from 1:50 to 7:11 UT on 2016 January 8, when we 
obtained 4650 and 4950 spectra in the red and blue
arms of ISIS, respectively. The difference is due to a technical problem
where two cubes in the red arm were not saved. This occurred around
2:23 and 4:01~UT where we lost approximately 20 minutes of red-arm observations.
For the subsequent analysis, we removed the first spectrum in each cube of 150
spectra, due to the difference in integration-time for the first frame compared to the remainder.
During the observations the seeing was around 1 arcsec or better. Conditions 
were photometric at least until 4:00 UT, based on a transparency plot from
the SuperWASP survey \citep{Pollacco16} from the same night. Afterwards, thin clouds
were likely around until about 5:30 UT, when there was an obvious cloud 
crossing for about 5 minutes. The guiding was lost for several seconds, but recovered
promptly. At 7:00 UT more clouds passed, which also coincided with
the end of the night (see Fig.~\ref{plot1}). Between 5:30 and 7:00 UT there
was likely to be more thin clouds present.    

\section{Data reduction}\label{reduction}

We used \textsc{iraf}\footnote{The Image
Reduction and Analysis Facility (\textsc{iraf}) is 
distributed by the National Optical Astronomy Observatories, which are
operated by the Association of Universities for Research in Astronomy, Inc., under
cooperative agreement with the National Science Foundation.} for bias subtraction
and wavelength calibration. Arc spectra were taken just before and after the 
observations. The wavelength solution was determined by 
a linear interpolation between the two sets of arcs. 
We used data without a flat-field correction, as this
provided light curves with a lower rms. The spectra
of WD~1145 and the reference star were slightly tilted. Because they
both have low count levels (their fluxes are only a few 
photo-electrons per wavelength step per frame), it is difficult to accurately fit
the trace of the spectra in each individual frame. 
If the spectral trace is constant in time, we could stack multiple frames to overcome this problem. 
Unfortunately, we found that positions of spectra were moving along the spatial direction with
timescales of roughly 1--2 minutes with a semi-amplitude of 2--3 pixels,
equivalent to about 0.5 arcsec. Therefore, we used \textsc{python} to implement the following data
reduction strategy. First, we determined a 2nd order polynomial trace and
profile for optimal extraction for both stars after stacking each cube of 150 images.
Then, we collapsed each individual image
within each cube in the dispersion direction, taking into account the spectral tilt, and
fitted a Gaussian function to both stellar profiles to get offsets in the
spatial direction with respect to the original trace. Finally, we extracted
each individual spectrum by summing all flux in the aperture 17 pixels
wide. We note that as we are using EMCCDs, this does not introduce extra
noise, as the readout noise is essentially zero. We used a trace corrected for offsets in the spatial 
direction. The sky regions for the background subtraction were shifted
by the same offsets and were located on both sides of a stellar aperture with
a size of 41 pixels each. To remove cosmic rays we used a median filter on
data with a size of 11 by 11 pixels, and set a level of cosmic rays to 1200
counts above median.

\section{Data analysis}\label{analysis}

\subsection{Photometric time series}\label{photometric}

We summed the flux over each extracted spectrum to obtain a photometric signal
for the target and a reference star. In the red arm, we constructed a
differential light curve by dividing the two time-series. In the blue arm,
however, this was not possible due to the reference star being too faint.
In Fig.~\ref{plot1} we show light curves
binned to an effective exposure time of 1 minute. The top panel of Fig.~\ref{plot1} 
shows the differential light curve for the red arm, and the bottom panel 
shows the raw light curve for the blue arm. Both light curves  
were normalized by fitting a 4th order polynomial to the
out-of-transit data. Photometric uncertainties were calculated as rms errors over
out-of-transit normalized light curves, using data binned in 1 minute in
time, and were assumed to be constant in time. 

\begin{figure*}
\includegraphics[width=\textwidth]{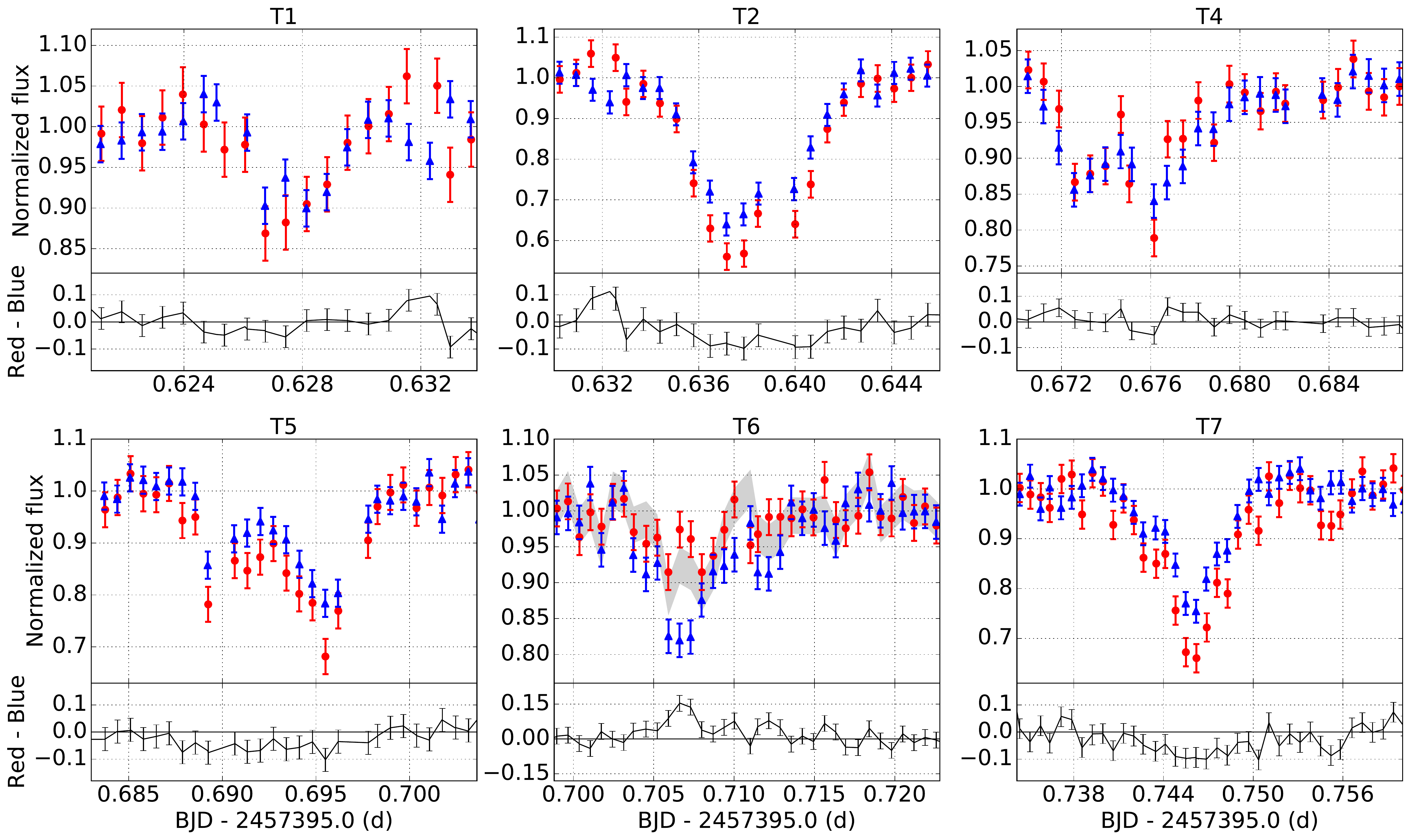}
\caption{For each transit T1--T7 upper panels show light curves observed during the night of 
2016 January 7 with the ISIS/QUCAM2 red arm (red points) and ISIS/QUCAM3 blue arm (blue
triangles). At lower panels for each transit we plot the difference of the differential red arm 
and raw blue arm light curve (black solid line) with corresponding error bars. For all
except for transit T6 we detect a colour difference of the order of 0.05--0.1 mag, 
with transits being deeper in the red arm. For transit T6 we show in addition 
at the upper panel a raw flux for the red arm (grey shaded region), as this
is the only transit for which we observe a difference between a differential
and raw flux in the red arm.
We normalized each transit by fitting a 1st order polynomial to the out-of-transit data 
immediately before and after the transit. Photometric uncertainties were 
calculated in a same way as in Fig.~\ref{plot1}, but using out-of-transit data 
immediately before and after each transit. BJD is Barycentric Julian Date.}
\label{plot2}
\end{figure*}

We observed six full transits, shown in Fig.~\ref{plot1}, and one
partial transit (T3), which unfortunately coincided with the technical 
issues in the red arm resulting in the loss of about 10 minutes of observations. 
The normalization of the entire (masked) light curve with a 4th order polynomial of time
was not optimal, particularly for the blue arm, and therefore we normalized 
each full transit separately by fitting a 1st order
polynomial to the out-of-transit data immediately before and after the
transit. These final, normalized light curves are shown in Fig.~\ref{plot2} along with
their differences for all six full transits.
Photometric uncertainties were calculated in the same way as before,
but only using out-of-transit data immediately before and after each transit.
The out-of-transit data used for a normalization and error calculations 
are displayed in Fig.~\ref{plot2}, and we assumed constant uncertainties
over each time interval. 

Our observations (except for transit T6)
show an evident colour difference between the in- and out-of-transit photometry, 
of the order of 0.05--0.1 mag, where transits are deeper in the red arm (see Fig.~\ref{plot2}). 
We calculated the significance level of this detection from the differences of the red
and blue light curve between the mid-ingress and mid-egress of each transit.
The resulting significance levels and colour differences are shown in Table~\ref{tab:sig_level}. 
We measured the most significant colour difference for transit T7 with 20$\,\sigma$
significance. Similar results were reported by \citet{Hallakoun17} who detected bluing 
with 2--3$\,\sigma$ significance, i.e. deeper transits in the redder bands 
({\it u}$^{\prime}-${\it r}$^{\prime}$ of up to $-0.05\jed{mag}$).

\begin{table} \tabcolsep=5pt
 \caption{The significance level of colour difference between the in- and
out-of-transits.}
\begin{center}
 \label{tab:sig_level}
 \begin{tabular}{lcccccc}
  \hline
  Transit & T1 & T2 & T4 & T5 & T6 & T7\\
  Red-Blue (mag) & -0.06 & -0.10 & -0.05 & -0.10 & +0.15 & -0.10\\
  Significance ($\sigma$) & >2 & 14 & 5 & 15 & 16--20 & 20\\
  \hline
 \end{tabular}
\end{center}
\end{table}

In our analysis, we compared a differential light curve for the
red arm (i.e. corrected with the reference star) and a raw light curve for the blue arm 
(i.e. with no reference star).
For a consistency check, we also compared raw light curves for both arms. 
For all transits except for transit T6, differences between using
differential and raw light curves for the red arm are insignificant, and so our results are 
also valid in this case. This is expected as the conditions were photometric during 
most of the night, and long term trends were removed by our analysis.

However, during transit T6, we observed the transit being deeper by about 0.15 mag 
in the blue arm compared to the differential flux in the red arm, with $20\,\sigma$ significance.
This is the only transit for which we observe any significant difference between
differential and raw fluxes in the red arm. In Fig.~\ref{plot2}, for transit T6
we also show the raw flux for the red arm (grey shaded region). 
If we compare raw fluxes in both arms transit T6 is still deeper in the blue arm, 
with a slightly lower significance of $16\,\sigma$. 
If this transit is real, it could be a sign of small dust grains in the transiting
material, however, a signal faded and a full width at half maximum of
stellar spectra degraded for a significant amount of time during transit T6,
and we suspect that a thin cloud was passing by our field. This
would also explain the difference between the differential and raw flux in the red
arm, which is consistent during the remainder of the observations.
We include transit T6 here as we detect a transit-like feature in the
differential light curve,
but conclude that this is most likely the result of
imperfections of the differential photometry when thin clouds are passing by.

\subsection{Spectroscopic time series}\label{spectroscopic}

\begin{figure}
\includegraphics[width=\columnwidth]{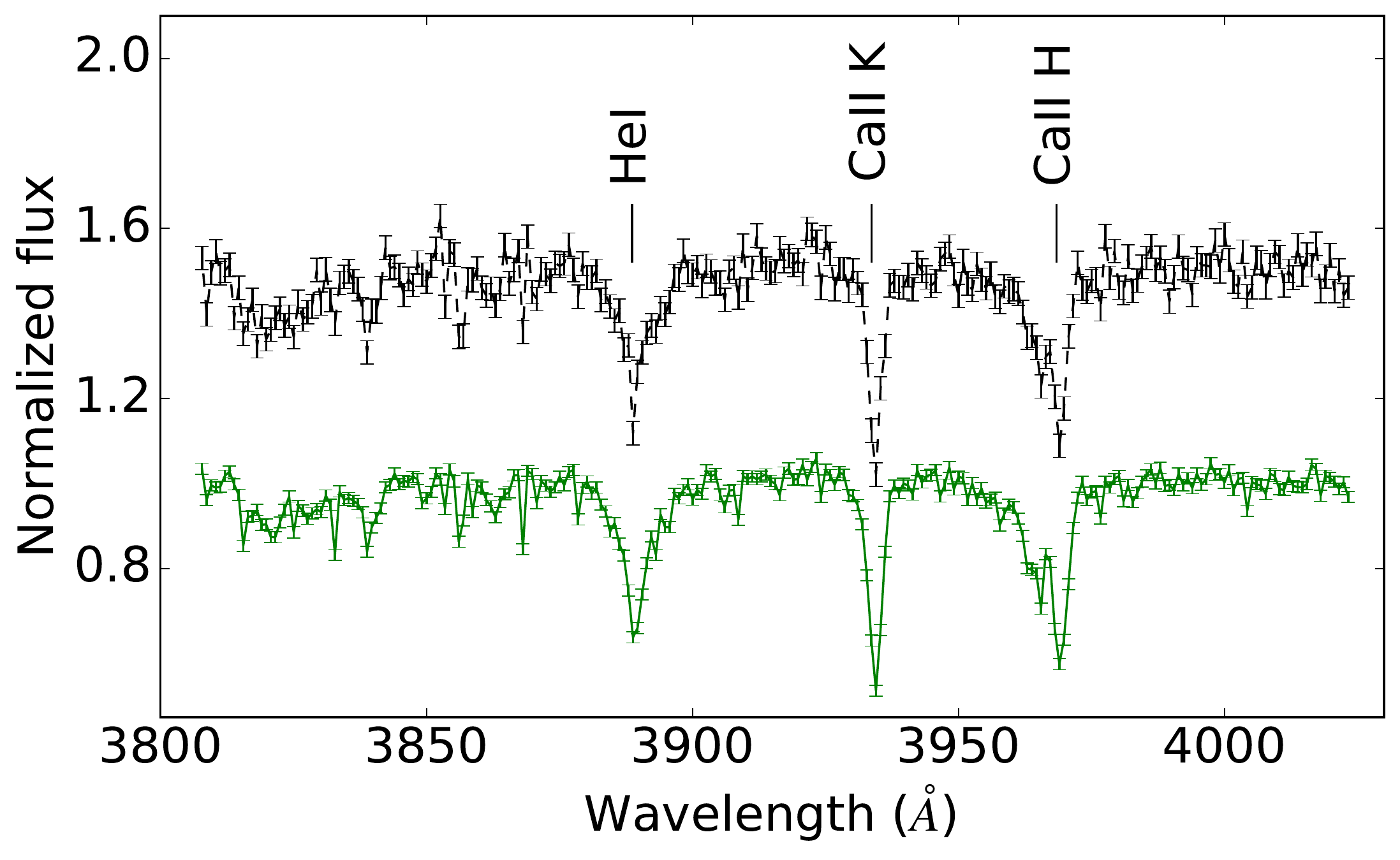}
\caption{Blue arm averages of all out-of-transit (green solid line, bottom) and in-transit
(black dashed line, top) spectra binned by 4 pixels in the spectral 
direction, plotted with corresponding error bars. 
The in-transit spectrum is shifted by 0.5 for clarity.} 
\label{plot3}
\end{figure}

\begin{figure}
\includegraphics[width=\columnwidth]{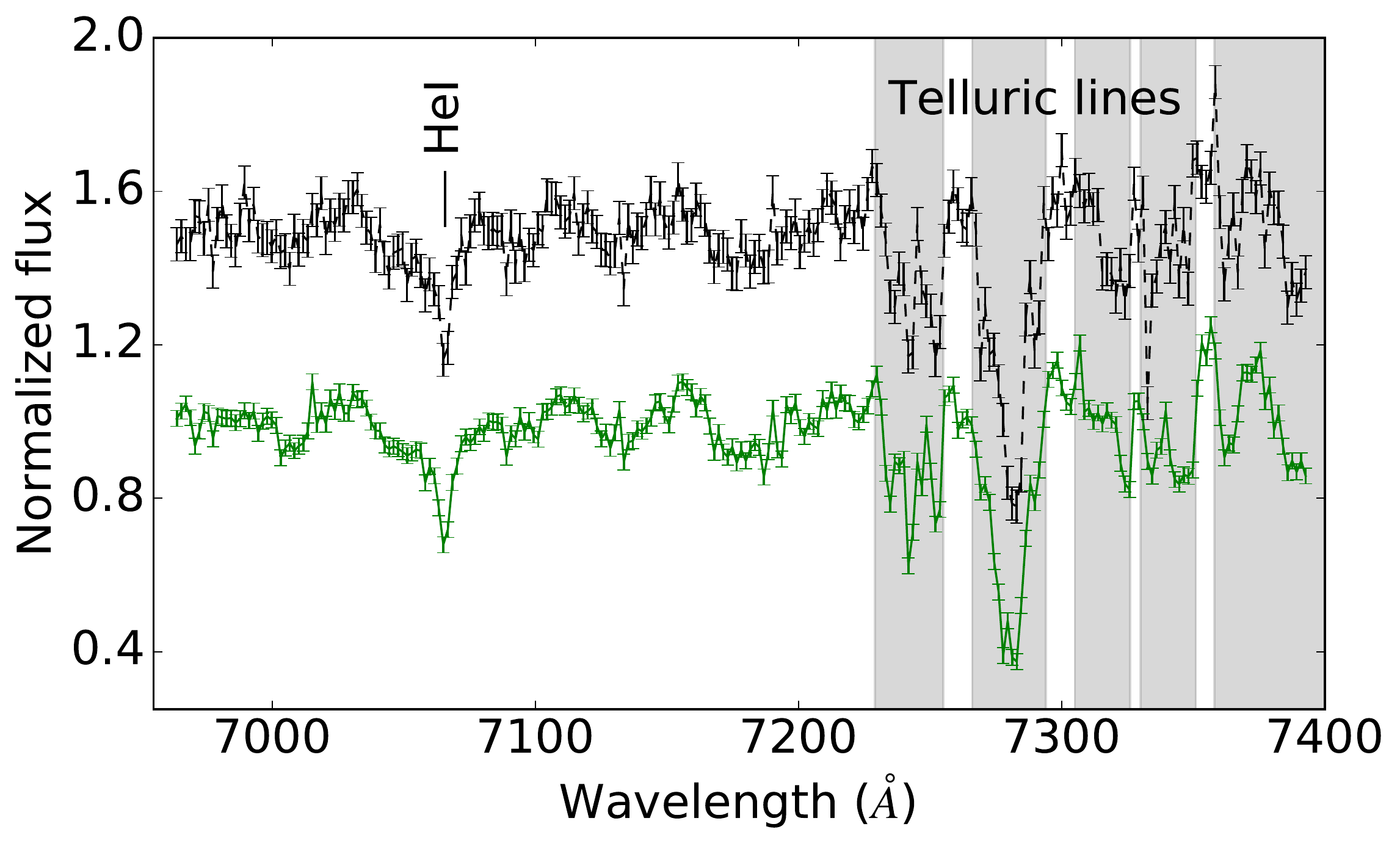}
\caption{Red arm averages of all out-of-transit (green solid line, bottom) and in-transit
(black dashed line, top) spectra binned by 4 pixels in the spectral 
direction, plotted with corresponding error bars. 
The in-transit spectrum is shifted by 0.5 for clarity.
Telluric lines are present in spectra and plotted with grey background areas.} 
\label{plot4}
\end{figure}

To normalize the data, we fitted a 6th order Legendre function in the continuum 
of all combined spectra and then divided each individual spectrum by this
function. Our observations lasted for more than 5
hours, and in order to correct for atmospheric refraction at higher
airmass, we fitted a second order polynomial to each
spectrum to correct for possible tilts between individual spectra.
From the red-arm differential light curve (see Fig.~\ref{plot1} top), we determined in-
and out-of-transit times. During the out-of-transit times, we averaged
spectra over 10 minute intervals, except for the last interval in each
out-of-transit range for which spectra were averaged over the remaining time. This was
typically more than 2.5 minutes, otherwise the continuum scatter was too
high and the data were discarded. During the in-transit times, we averaged spectra
over the whole transit duration (see Fig.~\ref{plot1}, green
background areas). We then binned all of the resulting spectra by 4
pixels in wavelength. We applied barycentric corrections to time and radial velocities 
using \textsc{BarCor}\footnote{http://www.ing.iac.es/$\sim$mh/}, 
and we adopted the system's radial velocity of $12$\,km\,s$^{-1}$
\citep{Xu16, Redfield17}. 
In Fig.~\ref{plot3} and Fig.~\ref{plot4} we show averages of all out-of-transit 
and in-transit spectra for the blue and red arms, respectively. The uncertainties were
calculated as a standard deviation of points out of the spectral lines 
over the wavelength and also over time, when we assumed that they are
constant in wavelength or time. Resulting errors of the two
approaches were comparable, and we used the former for the final interpretation of data.
In the subsequent analysis, we will focus on the following lines: 
\ion{He}{i} 3888.65~\AA, \ion{Ca}{ii} 3933.66~\AA~(K line) and \ion{Ca}{ii} 3968.47~\AA~(H line) in
the blue arm, and \ion{He}{i} 7065~\AA~line in the red arm, which is free
of contamination from telluric lines (see Fig.~\ref{plot4}).

We measured that the position of lines was changing during the
observations despite of earlier correction of the wavelength scale between
the beginning and end of time-series. This is most likely due to the atmospheric
differential refraction. In the blue arm, between the beginning 
and end of observations, \ion{He}{i} 3888.65~\AA~line changed the position by 0.40~\AA, 
\ion{Ca}{ii} 3933.66~\AA~line by 0.44~\AA~and \ion{Ca}{ii} 3968.47~\AA~line by 0.50~\AA.
Also, the observed spectral line wavelengths are redder than air wavelengths
for all three lines (see Table~\ref{tab:line_centers}). In the red arm, between the beginning and end of observations, 
\ion{He}{i} 7065~\AA~line changed the position by 0.80~\AA. 
The expected change of line positions between the beginning and end of observations due to the atmospheric 
differential refraction is approximately 0.44 and 0.68~\AA~for the blue and red arm,
respectively, which agrees well with our measurements. 

\begin{figure*}
\includegraphics[width=\textwidth]{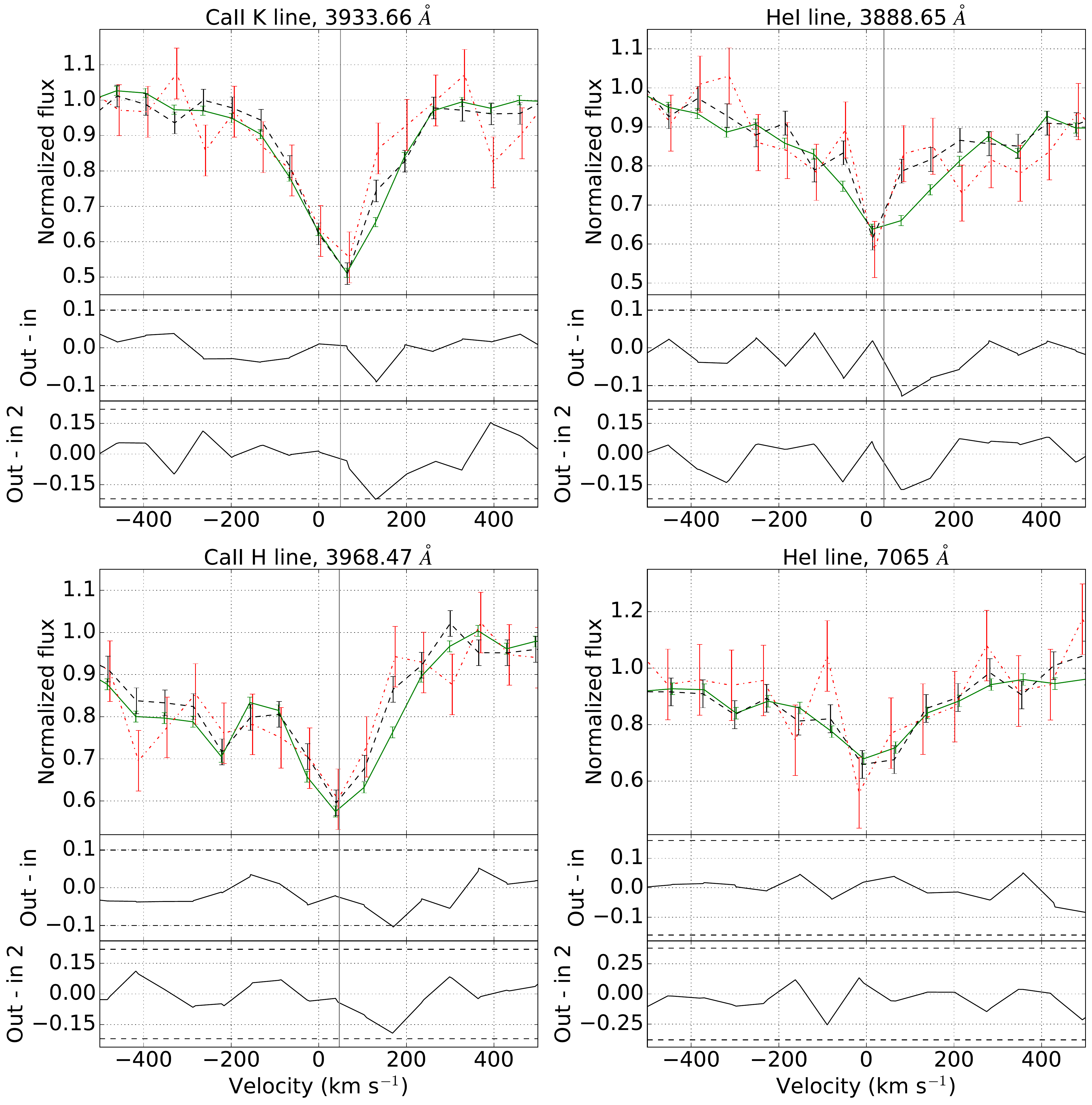}
\caption{For the three main detected lines in the blue arm and the \ion{He}{i} line in the red arm, top panels show 
average of all out-of-transit (green solid line) and all except transit T6 in-transit 
spectra (black dashed line). Also shown is the in-transit spectrum during transit
T2 (red dash-dotted line), which was the deepest transit during our
observations. Middle (bottom) panels show a difference between average out-of-transit
and in-transit spectrum (spectrum during transit T2) (black solid line).
The black dashed horizontal lines in the two lower panels
for each line indicate the $3\,\sigma$ significance level. The vertical grey solid line shows observed
wavelength for the blue lines.} 
\label{plot5}
\end{figure*}

To assess whether there is a difference in spectral lines during transits
compared to out-of-transit, for each line in the blue arm and the \ion{He}{i} line in the red arm 
we show in Fig.~\ref{plot5} 
in upper panels the average of all out-of-transit (green solid line) 
and all (except transit T6, which was excluded due to its speculative nature) 
in-transit spectra (black dashed line). We also
display in-transit spectrum during transit T2 (red dash-dotted line), which was the deepest
transit during our observations. For each line in Fig.~\ref{plot5} we plot
in middle panels the difference between average out-of-transit and in-transit
spectrum, and in bottom panels the difference between average out-of-transit
and in-transit T2 spectrum (black solid lines). The uncertainties were calculated as a standard 
deviation of points out of the spectral lines over the wavelength when we 
assumed that they are constant in wavelength.

For all three spectral lines in the blue arm, we detect with $>6\,\sigma$ significance
shallower lines during transits compared to the out-of-transit data (see
Fig.~\ref{plot5}). The significance level was calculated from the differences of the out-of-transit
and in-transit spectra across each line profile. It also appears that during transits
the reduction in absorption is larger on the red side 
of a spectral profile for the two circumstellar lines (\ion{Ca}{ii} K and H).  
This has been reported by \citet{Redfield17} who observed 
that the absorption on the red side of spectral profiles decreased during
transits on 2016 March 29 and 2016 April 08. They explored time-resolved spectra
and observed $>3\,\sigma$ variations from the mean profile in a significant
fraction (i.e., >10 per cent) of velocities across the circumstellar profile. 
Recently, \citet{Izquierdo18} unambiguously confirmed the equivalent 
width decrease of the circumstellar absorption lines during the deepest and longest transit 
in their observations, supporting spatial correlation between the circumstellar gas
and dust. Surprisingly, we also observe a reduction in absorption during transits for
the photospheric \ion{He}{i} 3888.65~\AA~line, which has a similar behaviour
as the two circumstellar lines in the red side of a spectral profile, but 
in addition has a decreased absorption in the blue side of a spectral profile.
For the \ion{He}{i} line in the red arm we do not observe any differences of
a spectral line profile between in-transit and out-of-transit spectra 
(see Fig.~\ref{plot5}). This could be explained by a higher excitation energy
of this line compared to the \ion{He}{i} 3888.65~\AA~line.

We note that our results are not caused by variable seeing during observations.
In order to check this, we made Fig.~\ref{plot5} using only out-of-transit data in the best
seeing, and the differences between in-transit and out-of-transit line shapes
were even larger. 

\begin{table}
\begin{center}
 \caption{Spectral lines air and observed wavelengths
$\lambda$.}
 \label{tab:line_centers}
 \begin{tabular}{lcccc}
  \hline
  sp. line & Air $\lambda$ & Observed
$\lambda$ & \multicolumn{2}{c}{difference}\\
  & \AA & \AA & \AA & km\,s$^{-1}$ \\
  \hline
  \ion{He}{i}  & 3888.65 & 3889.17 & 0.52 & 40\\
  \ion{Ca}{ii} & 3933.66 & 3934.31 & 0.65 & 50\\
  \ion{Ca}{ii} & 3968.47 & 3969.09 & 0.62 & 47\\
  \hline
 \end{tabular}
\end{center}
\end{table}

\section{Conclusions}\label{Conclusions}

We present time resolved spectroscopy of WD~1145, discovered to exhibit
transiting signals of orbiting material. The observations were taken on the
night of 2016 January 7 with electron-multiplying frame-transfer CCD cameras (QUCAMs)
\citep{Tulloch11}, in two different spectral ranges: the blue (from 3800--4025~\AA)
and the red (from 7000--7430~\AA). While the individual exposures have
a low signal-to-noise per spectral pixel (exposure times were 3.6 s), the lack of
readout noise with QUCAM CCDs means that we could choose the temporal binning of the 
light curve to optimize between signal-to-noise ratio, spectral resolution and time
resolution based on the observed transit depths and durations.

When comparing individual transits in two different spectral ranges, our
observations show with $20\,\sigma$ significance an evident colour difference between the in- and
out-of-transit data, of the order of 0.05--0.1 mag, where transits are deeper in the red arm. 
The bluing is surprising and not usual in dusty environments, typically showing reddening. 
Our observations thus confirm findings of \citet{Hallakoun17}, who detected bluing 
with 2--3$\,\sigma$ significance, of the order of {\it u}$^{\prime}-${\it r}$^{\prime}$ up
to $-0.05\jed{mag}$.

The analysis of spectral lines in the blue arm shows with $>6\,\sigma$ significance that
lines are shallower during transits compared to the out-of-transit data. It also appears that during transits
the reduction in absorption is larger on the red side of a spectral profile for the 
two circumstellar lines (\ion{Ca}{ii} K and H). This confirms results reported by \citet{Redfield17}  
who observed that the absorption on the red side of spectral profiles decreased during
transits on 2016 March 29 and 2016 April 08. They explored time-resolved spectra
and observed $>3\,\sigma$ variations from the mean profile in a significant
fraction (i.e., >10 per cent) of velocities across the circumstellar profile.
Recently, \citet{Izquierdo18} unambiguously confirmed the equivalent 
width decrease of the circumstellar absorption lines during the deepest and longest transit 
in their observations, supporting spatial correlation between the circumstellar gas
and dust. Surprisingly, we also observe a reduction in absorption during transits for
the photospheric \ion{He}{i} 3888.65~\AA~line, which has a similar behaviour
as the two circumstellar lines in the red side of a spectral profile, but 
in addition has a decreased absorption in the blue side of its spectral profile.
For the \ion{He}{i} line in the red arm we do not observe any differences of
a spectral line profile between in-transit and out-of-transit spectra, but
this can be explained by a higher excitation energy of this line compared 
to the \ion{He}{i} 3888.65~\AA~line.

Both detections of wavelength-dependent transits for WD~1145 (this analysis and
\citet{Hallakoun17}) are the only ones so far which included observations in the
{\it u}$^{\prime}$-band range, 
where WD~1145 displays many broad absorption lines caused by circumstellar gas.
The fact that other studies \citep{Alonso16, Zhou16, Croll17,
Izquierdo18, Xu18} did not observe wavelengths shorter
than 4300~\AA~could explain why they did not detect bluing. 
As discussed by \citet{Hallakoun17} and \citet{Redfield17}, the most
plausible explanation for the {\it u}$^{\prime}$-band excess during transits is the reduced
circumstellar absorption along the line-of-sight. As suggested by
\citet{Redfield17}, the decrease in line absorption during transits can be explained 
if an opaque body, like a parent body or it's fragments, blocks a fraction of the gas 
disc causing the absorption, which means that the absorbing gas is between
the white dwarf and the transiting objects.  

For a consistency check of our results, we also compared raw light curves for
both arms. We found that for all transits except for transit T6, differences between 
using the differential and raw light curve for the red arm are insignificant. 
During transit T6, we observe the transit being deeper by approximately 0.15 mag 
in the blue arm compared to the differential flux in the red arm, with $20\,\sigma$ significance.
This is the only transit for which we observe significant difference between
differential and raw fluxes in the red arm. If we compare raw fluxes in both arms transit 
T6 is still deeper in the blue arm, with less significance of $16\,\sigma$. 
If this transit is real, it could be a sign of small dust grains in the transiting
material. However, we suspect that a thin cloud was passing by our field
during this time, which explains the difference between the differential and 
raw flux in the red arm. We conclude that transit T6 is most likely not to be a real transit event,
and is caused by thin clouds passing by.

Finally, our results demonstrate the capability of QUCAM CCDs at the 4.2-m
William Herschel Telescope (La Palma, Spain) to perform high-quality time resolved 
spectroscopy of relatively faint targets (r = 17.3 mag). \citet{Rappaport18}
reported that the optical activity of WD~1145 has increased to the highest level observed since
its discovery, with approximately 17 per cent of the optical flux extinguished per
orbit, and some transits with depths of up to 55 per cent and durations as long as
two hours. As the transiting source material continues to evolve with time, 
there is likely to be a continuous deposition of material feeding the disc of gas responsible for the
circumstellar absorption \citep{Redfield17}. Future observations of WD~1145
with QUCAMs would be helpful to reveal changes of the circumstellar line profiles which would
aid in understanding the geometry of the innermost circumstellar gas and
provide constraints on white dwarf disc accretion models \citep{Cauley18}.  
In addition, it would be interesting to re-observe the system during high
activity levels to confirm a reduction in absorption during transits for
the photospheric \ion{He}{i} 3888.65~\AA~line. 

\section*{Acknowledgements}

We would like to thank the referee for helpful feedback.
N.~P.~G. gratefully acknowledges support from the Royal Society 
in the form of a University Research Fellowship.
This research has made use of the electronic bibliography maintained by NASA-ADS
system, the SIMBAD database, operated at CDS, Strasbourg, France, and the VALD database, 
operated at Uppsala University, the Institute of Astronomy RAS in Moscow, and the University of Vienna.




\bibliographystyle{mnras}
\bibliography{references} 





\bsp	
\label{lastpage}
\end{document}